\documentclass[aps,pra,twocolumn,superscriptaddress]{revtex4-1}
\usepackage{graphics}
\usepackage{mathtools}
\usepackage{float}
\usepackage{bbold}

\usepackage{soul,xcolor}
\usepackage{graphicx}
\usepackage{amsmath,amssymb,amsfonts}
\usepackage{xcolor}
\usepackage{soul,xcolor}
\usepackage{tikz,tkz-euclide}
\def\DJo{$\;$\kern-.4em \hbox{D\kern-.8em\raise.15ex\hbox{--}\kern.35em okovi\'c}}
\def\CC{{\rm\kern.24em \vrule width.04em height1.46ex depth-.07ex
\kern-.30em C}}
\def\P{{\rm I\kern-.25em P}}
\def\NN{{\rm I\kern-.15em N}}
\def\RR{{\rm
         \vrule width.04em height1.57ex depth-.0ex
         \kern-.03em R}}
\def\RR{{\rm I\kern-.23em R}}
\def\id{{\rm 1\kern-.22em l}}
\def\ZZ{{\sf Z\kern-.44em Z}}

\newtheorem{psatz}{Satz}[section]
\newtheorem{pdef}{Definition}[section]
\newtheorem{conjecture}{Vermutung}[section]

\newenvironment{eqblock}[2]{\beq\label{#2}\begin{array}{#1}}{\end{array}
                                \eeq}
\newenvironment{neqblock}[1]{\[\begin{array}{#1}}{\end{array}\]}
\newcommand{\beqb}{\begin{eqblock}}
\newcommand{\eeqb}{\end{eqblock}} 
\newcommand{\nbeqb}{\begin{neqblock}}
\newcommand{\neeqb}{\end{neqblock}}

\newcommand{\eps}{\varepsilon}
\newcommand{\beq}{\begin{equation}}
\newcommand{\beqa}{\begin{eqnarray}}
\newcommand{\eeq}{\end{equation}}
\newcommand{\eeqa}{\end{eqnarray}}
\newcommand{\nbeqa}{\begin{eqnarray*}}
\newcommand{\neeqa}{\end{eqnarray*}}

\newcommand{\bra}[1]{\langle #1 |}
\newcommand{\ket}[1]{| #1 \rangle}
\newcommand{\ketbra}[1]{\ensuremath{| #1 \rangle \langle #1 |}}

\newcommand{\Matrix}[2]{\left( \begin{array}{#1} #2 \end{array}
  \right)}

\usepackage{graphicx} 
\usepackage[colorlinks=true,citecolor=blue,linkcolor=blue,urlcolor=blue]{hyperref}
\begin{document}

\title{Threetangle in the XY-model class with a non-integrable field background}

\begin{abstract}
The square root of the threetangle is calculated for the transverse XY-model with an integrability-breaking in-plane field component. To be in a regime of quasi-solvability of the convex roof, we concentrate on a 4-site model Hamiltonian. In general, the field and hence a mixing of the odd/even sectors, has a detrimental effect on the threetangle, as expected. Only in a particular spot of models with no or weak inhomogeneity $\gamma$ does a finite value of the tangle prevail in a broad maximum region of the field strength $h\approx 0.3\pm 0.1$. There, the threetangle is basically independent of the non-zero angle $\alpha$. This system could be experimentally used as a quasi-pure source of threetangled states or as an entanglement triggered switch depending on the experimental error in the field orientation.
\end{abstract}
\date{November 2024}
\author{J\"org Neveling}
\affiliation{Software Engineering, Konecranes GmbH, Forststrasse 16, 40597 D\"usseldorf, Germany.}
\author{Andreas Osterloh}
\affiliation{Quantum Research Center, Technology Innovation Institute, Abu Dhabi, P.O. Box 9639, UAE}

\maketitle

{\em Introduction --}
Entanglement is one of the main ingredients of modern quantum technology applications like quantum computation\cite{ladd2010quantum,gill2022quantum,orus2019quantum,ringbauer2022universal}, quantum sensing and metrology\cite{narducci2022advances,degen2017quantum,crawford2021quantum,wang2023photonic,aslam2023quantum,schnabel2010quantum,pezze2018quantum,polino2020photonic}, and quantum communication\cite{gisin2007quantum,luo2023recent,cozzolino2019high,de2024parallel}. However, finite temperature and experimental imperfections cast theoretically ideal models to the experimentally realistic realm of feasibility.
Therefore it is extremely important to analyse the effect these imperfections have on this valuable resource. 
The focus here lies on multipartite entanglement\cite{horodecki2009quantum,guhne2009entanglement,eltschka2014quantifying,hein2004multiparty} as a resource. There are multiple ways to measure whether some multipartite entanglement might be in the system. The so-called Genuine Multipartite Entanglement\cite{toth2005detecting,huber2010detection,huber2014witnessing,milz2021genuine} and the Generalized Multipartite Negativity\cite{vidal2002computable,jungnitsch2011taming}
deliver useful information as far as a coarse grained classification of the entanglement content beyond bipartite entanglement is concerned. These are forged from bipartite measures of entanglement, as the Partial Transpose, and (linear) entropies of entanglement and are constructed to be zero for convex combinations of bipartite states. Another philosophy is applied in the (Generalized) Geometric Entanglement\cite{wei2003geometric,sen2010channel,das2016generalized,kondra2021stochastic,blasone2008hierarchies},
where the distance to a certain class of states is detected. These can be, but do not necessarily need to be, the separable states. For being more selective on the special type of entanglement, there exist measures for pure states\cite{VerstraeteDM03,OS04,osterloh2016exact}, which are sufficiently relevant due to the vast amount of interesting features they accumulate in a single concept\cite{OS04,DoOs08,osterloh2010invariant}. To clearly distinguish from its merely non-bipartite cousin, we rename it into {\em genuine multipartite SL-entanglement}. We preserve the term {\em tangle} as an umbrella term for multipartite SL-entanglement measures. The drawback of these SL-invariant measures is that they have a non-linearity of $2n$ in the wave function entries. This lifts their convex-roof extension from the integrable regime for the concurrence with second degree non-linearity to being a highly non-trivial task \cite{gurvits2003classical}. A degree two measure of SL-entanglement still exist for all even number of qubits that is integrable with respect to obtaining their convex-roof but it only covers the generalized GHZ state\cite{osterloh2017four}.
Even though it is futile to look for an exact solution of the convex roof due to the NP-hardness of the problem\cite{gurvits2003classical} in general, it is extremely useful to study the behavior of optimal decompositions in simple cases, in order to elaborate an approximative scheme of structural similarity to perturbation theory along the lines of 
~\cite{osterloh2016upper,osterloh2016exact}.
We compare our results to a lower bound to the one parameter family of GHZ-symmetric states\cite{eltschka2014practical,eltschka2014quantifying}.
Obtaining new such lower bounds is cumbersome since it requires multiple parameters to be optimized over. 

We here study a quantity that singles out GHZ-type of SL-entanglement, namely the threetangle\cite{Coffman00} and analyze its dependence on an integrability breaking field in the class of transverse XY models. We utilize an upper bound approach in that we construct quasi-optimal decompositions of the density matrix in consideration,
but argue that the result is the exact convex roof because the maximal amount of states in this case is limited by $4$, and based on the state-locking\cite{osterloh2016exact} of the zero-polytope in optimal decompositions together with the non-intersection property of such optimal decompositions. 

\section{Model Hamiltonian}
We deal with the Hamiltonian
\beqa\label{eq:Ham}
H&=&-\sum_{j=1}^L\left[ \frac{1+\gamma}{2}\sigma_j^x\sigma_{j+1}^x+
\frac{1-\gamma}{2}\sigma_j^y\sigma_{j+1}^y \nonumber \right.\\
& & \quad \left. +h (\sigma_j^z \cos{\alpha}+\sigma_j^x\sin{\alpha} )\right]\; ,
\eeqa
where $L$ is the number of sites, $\gamma$ is the isotropy parameter, $h$ is the magnetic field, and $\sigma^{.}_j$ are the Pauli matrices.
For the angle $\alpha=0$ this is the integrable XY model in transverse field\cite{barouch1970statistical,barouch1971statistical2}. A non-zero angle $\alpha$ leads to a $\ZZ_2$ symmetry breaking, and thus the odd and even sectors, which before were conserved, become intertwined.
A non-zero $\alpha$ can be either due to experimental imperfections or it may be introduced willingly to destroy the integrability. Of course, it is interesting how various correlations behave and in particular would it be intriguing to see how entanglement quantifiers behave in presence of a non-zero angle.
Here, we are interested in quantifiers, which detect the genuine SL-invariant part of the entanglement content; among them, most prominently, the concurrence\cite{Wootters98} and the threetangle\cite{Coffman00} $\tau_3$ (see Appendix~\ref{app:threetangle}) which will be detected by $\sqrt{\tau_3}$\cite{viehmann2012rescaling}.
For the purposes of this work, we deal only with the four site model as it has the advantage of exact predictive power where the entanglement vanishes.
We draw a comparison with the lower bound from GHZ-symmetric states.

For a general $\alpha\neq 0$ the eigenstates of the Hamiltonian \eqref{eq:Ham} are superpositions of eigenstates of both parity sectors. Only the two groundstates will have an essential participation 
and the ground state will be approximately given by a superposition
$\ket{\psi_{\rm GS}}(p):=\sqrt{p}\ket{\psi_o}+\sqrt{1-p}\ket{\psi_e}$, $p\in\{0,1\}$, and $\partial_p H=0$, $\partial_p^2 H>0$.
However, for the four site model the ground state is found by means of an exact diagonalisation.
\begin{figure}[h!]
    \centering
    \includegraphics[width=0.95\linewidth]{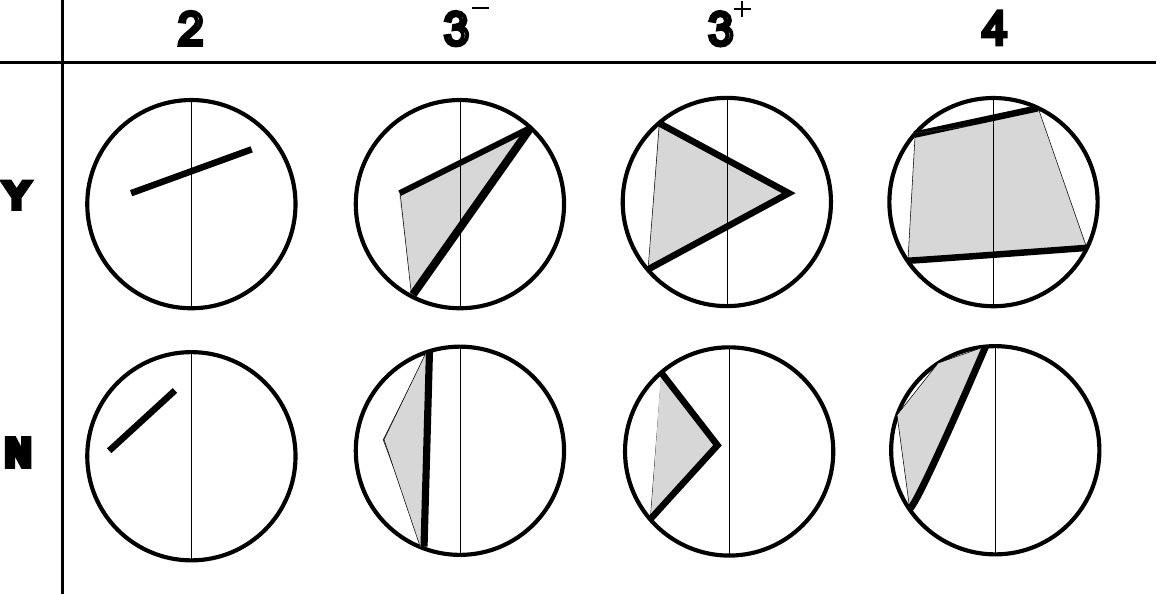}
   \caption{Complete classification of polytopes for real eigenstates of the density matrix in the real $x$-$z$ plane. Shown are typical examples for the model Hamiltonian. Every rotation of the zero polytope about the $y$-axis may lead to another classification. For the rightmost two columns, each of the distributions of $n$ pure states is possible with $n\in\{0,1,2,3 \}$ for both orientations.
   Here only the evenly distributed cases are depicted.} 
    \label{fig:zero-polytope-class}
\end{figure}
Given that the wavefunction for the groundstates is strictly real, 
the solutions for the zero-polytope~\cite{KENNLINIE,osterloh2016exact} are grouped in complex conjugated pairs. There are eight different types of polytopes of particular relevance here; for each of them
there is one intersecting the $z$-axis of the Bloch sphere (Y) and one that does not (N). They are shown in Fig.~\ref{fig:zero-polytope-class} in the ($x-z$) plane so that each complex solution $z$ and $z^*$ are projected onto the same point.
Therefore, each vertex in the inner part of this projected Bloch sphere represents two complex conjugated solutions. 
Each bold line marks a side of the polytope, which is visible from a state $\rho$ the $z$-axis and that therefore can be part of an optimal decomposition\cite{osterloh2016exact,Neveling-thesis} for it. For a point $\vec{b}$ inside this projection
of the Bloch sphere being member of an optimal decomposition for a density matrix $\rho_0$ lying on the $z$-axis at $\vec{r}_0$ of the Bloch sphere, it is clear that the remaining states from the zero-polytope that completes the optimal decomposition will correspond to a point $\vec{r}_1$ on the line $\lambda(\vec{b}-\vec{r}_0)$. 
Due to the reality of the ground state, we further have that every element $\vec{b}$ in the optimal decomposition has its complex conjugated solution $\vec{b}^*$ yielding the same value for the tangle. So, without loss of generality, we have that $\vec{b}_0=(\vec{b}+\vec{b}^*)/2$ corresponds precisely to the real point within the Bloch sphere projection. The line in between 
$\vec{b}$ and $\vec{b}^*$ must not intersect another optimal polytope; but since both points will correspond to states from the zero-polytope, this condition is trivially satisfied. Combined, both arguments lead to the real point $\vec{r}_1$ being part of the optimal decomposition that lies on the connecting line $\lambda(\vec{b}_0-\vec{r}_0)$.

Every decomposition is classified by two numbers, $n_0$ and $n_e$, by $(n_0,n_e)$. Whereas $n_0$ is the number of pure states taken from the zero-polytope, $n_e$ is the corresponding number of entangled pure states in the decomposition. 
Its tangle is given by convex combination of the corresponding density matrix: $\rho_0=\sum_{i=1}^{n_0} m_{i} \ketbra{Z_{\nu_i}} +\sum_{\eps=1}^{n_e} M_{\eps} \ketbra{T_{\eps}}$ $\Rightarrow \tau=\sum_\eps M_\eps \tau[T_{\eps}]$, where $Z_i$ are the vertices of the zero-polytope, $T_{\eps}$ are the entangled states in the decomposition and $\tau$ is the tangle.
Every decomposition that reaches the minimum for the mixed state $\rho_0$ is an optimal decomposition.
The main working assumption is that optimal decomposition must continuously vary in the available parameters. The probability $p$ connecting both eigenstates of the density matrix is the only parameter here. 

Our approach is to first look at the minimum solution with one single pure state out of the zero-polytope, i.e. optimal $(n_0,1)$-solutions. This gives some insight of how these optimal decompositions behave. We make use of the behavior known for optimal decompositions\cite{OS04,EOSU,osterloh2016upper,osterloh2016exact,viehmann2012rescaling}. In the next step we tried to look whether $(n_0,2)$-decompositions optimize the result, admitting the pure state to split into two, maintaining the connecting axis the optimal $(n_0,1)$ decomposition had with the density matrix. \\
The first non-trivial result is that, what we call "brachiating states", are optimal $(n_0,1)$-decomposition. For this, we have analysed polytopes of the type $3^+Y/N$ as the main contributors for this model (see Fig.~\ref{fig:zero-polytope-class}).
We list some general requirements on optimal decompositions in Appendix~\ref{app:optimal-decompositions}.

\section{Brachiating states}
Before turning to the more relevant case of two-dimensional zero-polytopes, we briefly want to highlight one-dimensional polytopes with two equally degenerate solutions. This is the integrable situation for the convex roof as for the concurrence: here every $(n,1)$ decomposition is optimal, giving the same result for the tangle. 
This case has been analyzed in Ref.~\cite{regula2016geometric,regula2016entanglement} and it could be observed also in this work.
\\
The symmetric case for type $4Y$ with two values $p_i$ with opposing phases is the only one occurring in the transverse models. The exact convex roof is obtained from a standard procedure, with corresponding pure states located precisely in the poles of the sphere as for the symmetric GHZ-W mixture\cite{LOSU}.
Interesting is the asymmetric case for which a case study is shown in Ref.~\ref{app:case-studies-4Y}.

Polytopes of type $2N$ occured for the XX model with non-orthogonal field. In these cases we have run an extra analysis of general $(2,1)$ decompositions for optimality due to an initial error in the algorithm. Both $(2,1)$ decompositions have been analyzed with the states  of the zero-polytope corresponding to the real parts $z_{0;i}$, $i=1,2$. It is found that the respective minimum is either in the middle or at the end points, hence for $(1,1)$ decompositions. The minimal tangle was found in the center $z_{0;i}$ which minimizes the distance to the mixed state in consideration; thus, it maximizes the corresponding weight on the zero-states. We want to stress that a minimum elsewhere on the Bloch sphere would indicate that a state with complex part would be singled out that forms an optimal $(2,1)$ decomposition; i.e., the same would be true for the corresponding complex conjugated images of these states. This would enlarge the optimal decomposition to $(2,2)$ type. This has never been found optimal in the model of consideration.

Both zero-polytopes $3^+w$ for $w=Y,N$, among others, do occur for certain regions of the parameters of \eqref{eq:Ham} and give interesting insights in the behavior of optimal decompositions.
First we discuss the $3^+N$ class which does not cross the central axis of the Bloch sphere. It is clear that the corresponding tangle will be non-zero outside the zero-polytope, hence for the whole polar axis of the Bloch sphere.
The two sets $3^\pm Y/N$ are similar in what the optimal decomposition is concerned; they differ in the orientation of the density matrix within the Bloch sphere. We describe the analysis of $3^+Y$ in more detail in Appendix~\ref{app:case-studies-3pmN}, but give a summary of the findings here.

\begin{figure}[h!]
    \centering
    \includegraphics[width=0.45\linewidth]{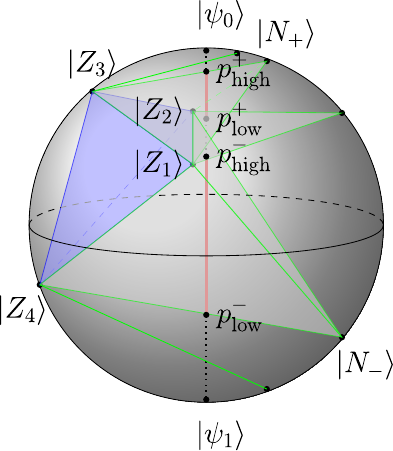}\hfill
    \includegraphics[width=0.45\linewidth]{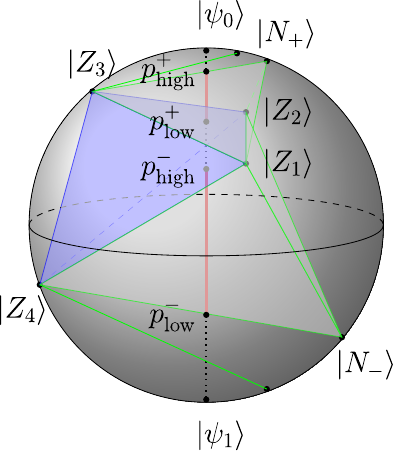}
   \caption{Polytopes of type $3^+N$ (left figure) and $3^+Y$ (right figure). Shown is that to the zero polytope (blue) a single pure state on the Bloch sphere is added corresponding to each of its zero-facets, respectively. They give rise to another simplex (grey triangular tetrahedra) lying on top of the corresponding 2-dimensional simplices of the zero-polytope. This is a general feature for every simplex on the surface of zero-polytopes. 
 Optimal decompostions, here indicated by green lines, brachiate through this structure of polytopes.}
    \label{fig:zero-polytope-sols}
\end{figure}

The relatively optimal decomposition of type $(n_0,1)$ can be obtained in two steps which is indicated in the right Bloch sphere in Fig.~\ref{fig:zero-polytope-sols}:
identifying the two points $p^+_{\rm low}$ and $p^-_{\rm high}$ where the z-axis leaves the zero polytope and the both points $p^+_{\rm high}$ and $p^-_{\rm low}$ where the z-axis leaves the respective $(3,1)$ polytopes with singled out states $\ket{N_\pm}$. These polytopes are attached to all the sides of the zero polytopes. 
The two lines $\overline{p^\pm_{\rm low}p^\pm_{\rm high}}$ where the tangle as a convex combination grows linearly\cite{LOSU,KENNLINIE,osterloh2016exact} 
are highlighted in bold. Beyond these linear behaviors the tangle behaves strictly convex as only a single zero state ($\ket{Z_3}$ or $\ket{Z_4}$) is left for optimal $(1,1)$ decompositions as indicated by green lines.

In the case of $3^+N$ polytopes,
we have to analyse both $(1,1)$ decompositions with $\ket{Z_{3}}$ and $\ket{Z_{4}}$ in comparison with the $(2,1)$-decomposition including the both states $\ket{Z_{1/2}}$. 
Here we focus on the two pure states $\ket{N_\pm}$ corresponding to the two facets visible from the $\psi_0$-$\psi_1$ axis. 
Due to the reality of the wavefunction, $\ket{N_\pm}$ come to lie at phases $\phi=0$ (the right meridian drawn) or $\phi=\pi$ on the Bloch sphere. 
In Fig.~\ref{fig:(3,1)-decomposition3+} it is seen that the minimum of these three curves is not convex. The single pure states $\ket{N_\pm}$ in a $(3,1)$-decomposition are product of the convexification procedure. This is demonstrated by the coinciding $p$-values of either state\footnote{In principle, it could be that different pure states were responsible for both convexification points. This would correspond to $5$ pure states in an optimal decomposition and their convex set forms the polytope. Though, always four out of the five states can be chosen as optimal decomposition. However, if the smallest curves of $(1,1)$ and $(2,1)$ decompositions are strictly convex in the range, this could never happen. This was always the case here.} in the right panel of Fig.~\ref{fig:(3,1)-decomposition3+}. Included is a search for optimality of $2$ entangled states in the decomposition (see Fig.~\ref{fig:intersectinglines} in appendix). The values of the tangle are shown exemplarily (green curve in Fig.~\ref{fig:(3,1)-decomposition3+}). A more detailed analysis of these deviations into both two dimensional $(2;2)$- or $(1;2)$-decompositions was never optimal for this Hamiltonian. However, it must be stated, that this is done by inspection, and a strict proof of this statement is missing. From first principles, we cannot say that these 2-dimensional decompositions are never optimal somewhere in the Bloch sphere. But they are penalized by the increase in weight of the entangled side.
\begin{figure}[h!]
    \centering
    \includegraphics[width=0.49\linewidth]{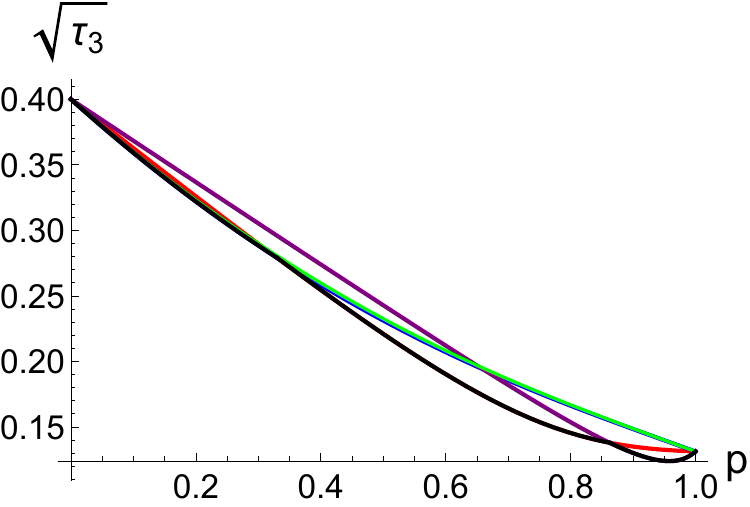}\hfill
    \includegraphics[width=0.49\linewidth]{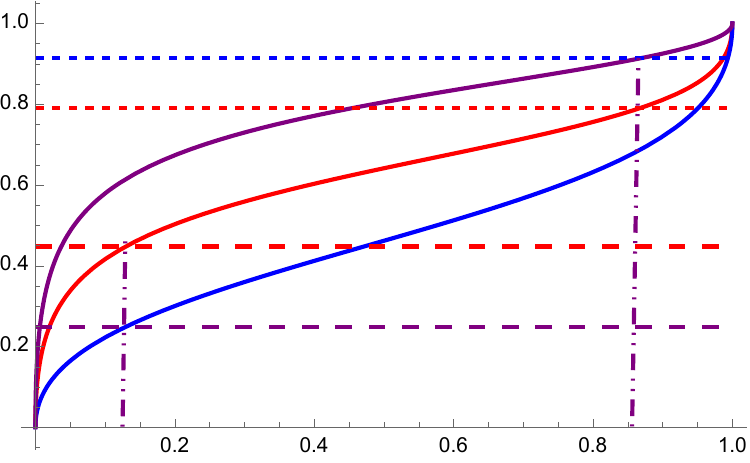}   \caption{$\sqrt{\tau_3}$ is shown in the left panel for the $(2,1)$  (red) and $(1,1)$ (blue and purple) decompositions for a zero-polytope of classes $3^+N$. The minimum is given by the black curve underneath. Convexifications are obviously needed. 
   It leads to a single 3D simplex on top of each visible sides of the zero-polytope as a basis. This is shown in the right panel where 
   $p_0$ of $\rho[p_0]$
   is plotted as a function of the $p$-value of the corresponding pure state as tip of that tetrahedron.
   }
    \label{fig:(3,1)-decomposition3+}
\end{figure}
Summarizing, a single pure state on the Bloch sphere is added to the zero polytope (blue) corresponding to each of its zero-facets. Together with the three pure states of the zero-polytope lying on top of the corresponding 2-dimensional simplices of the zero-polytope, they form three-dimensional polytopes outside of the zero polytope, highlighted in grey in Fig.~\ref{fig:zero-polytope-sols}. We want to stress that this is a more general feature for every simplex on the surface of zero-polytopes for general tangles of degree $2n$. 

The case of $3^-N$ polytopes is discusssed and a corresponding case study is shown in Fig.~\ref{fig:(3,1)-decomposition} of appendix~\ref{app:case-studies-3pmN}.

The optimal decompostions, indicated by green lines
in Fig.~\ref{fig:zero-polytope-sols}, brachiate through this structure of tetrahedra with the additional pure states for $\phi=0$. The optimal $(1,1)$-decomposition starts with the lowest pure state $\ket{Z_4}$ of the zero polytope and the second pure state brachiates through from $\ket{\psi_1}$ until it reaches $\vec{n}_- \mathrel{\widehat{=}}\ket{N_-}$ with  value $p_{\rm low}^-$ in $\rho(p_{\rm low}^-)$. This state gets fixed next in the lower grey pyramid and the state brachiates on the bisecting line in a $(3,1)$-decomposition until $\rho(p_{\rm high}^-)$ is reached, where it will be located in the center of the both states $\ket{Z_1}$ and $\ket{Z_2}$. Next these two states of the zero polytope get locked with a variable pure state on the Bloch sphere in a $(2,1)$-decomposition until it reaches $\ket{N_+}$ and $\rho(p_{\rm low}^+)$. Being this new pure state $\ket{N_+}$ the fixed point, the variable mixed state again brachiates along the bisecting line until it reaches the state $\ket{Z_3}$ in $(3,1)$-decompositions. Afterwards the optimal decomposition swings up in a $(1,1)$-decomposition until the upper pole of the Bloch-sphere is reached. This brachiating behavior of optimal decompositions was observed in Ref.~\cite{Neveling-thesis}.
It gives a scheme how optimal decompositions behave in general for the threetangle and in general for arbitrary tangles {\em iff} the density matrix has rank two.

\section{Results for non-zero alpha}
\begin{figure}[h!]
    \centering
    \includegraphics[width=0.49\linewidth]{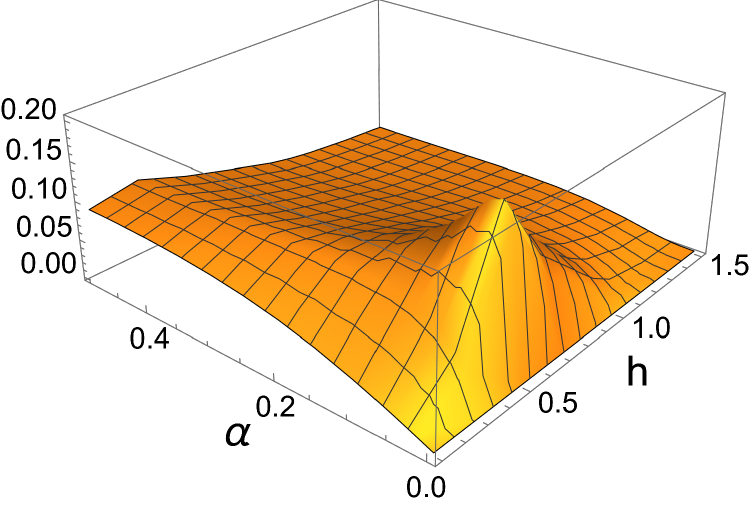}
    \includegraphics[width=0.49\linewidth]{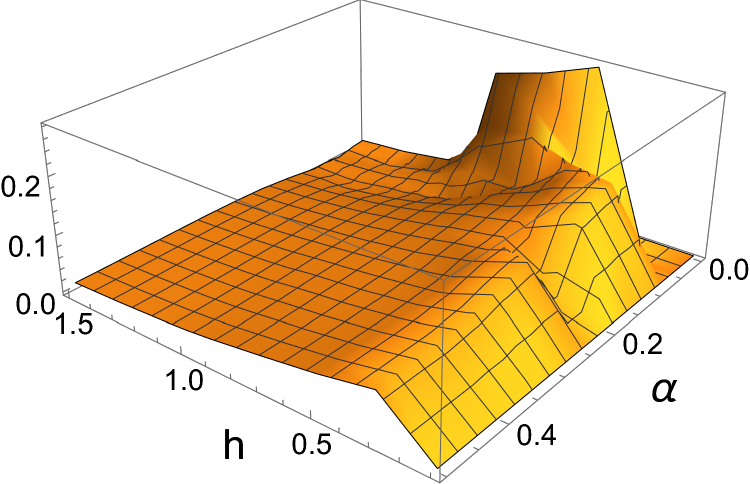}
    \caption{Entanglement detected by $\sqrt{\tau_3}$ for the reduced density matrix
    interpolating between both of its eigenstates as a function of the deviation angle $\alpha$ of the magnetic field. We show curves for $\gamma=0.0$ and $0.1$. The tangle $\sqrt{\tau_3}$ is algebraically decreasing with the angle $\alpha$. At about $h=0.3$ lies a broad maximum of $\sqrt{\tau_3}$ which is not notably changed in varying $\alpha$. From $\gamma=0.5$ (not shown) this effect is absent and the tangle in the system is destroyed immediately. Therefore the field to be transversally aligned is a hot-spot which does not support non-transversality of the field.
    }
    \label{fig:tau3-XY-nontransverse-small-gamma}
\end{figure}

For the non-transverse XY-model, we observe that its largest eigenvalues for three-site density matrices is very close to one such that $(1;1)$- and $(2;1)$-decompositions are optimal. We nowhere observed that the value for $p$ of the model considered was inside a convexified area.
The results are demonstrated in Fig.~\ref{fig:tau3-XY-nontransverse-small-gamma}, where $\alpha$ is shown in multiples of $\pi$ from $\alpha=0$ to $\alpha=\pi/2$. For larger values of $\gamma$ the value for $\sqrt{\tau_3}$ quickly decays for $\alpha\neq 0$, as will be discussed in a forthcoming paper. We show curves for $\gamma=0.0$ and $0.1$.  
$\sqrt{\tau_3}$ is algebraically decreasing with the angle $\alpha$
For the XX model there is a pronounced peak at finite $\alpha$ with a maximum roughly at $h=0.5$ and $\alpha=0.03$;
while $\sqrt{\tau_3}$ decays algebraically from this maximum to a relative minimal value at $\alpha=\pi/2$, $\sqrt{\tau_3}$ proves to be essentially unaffected by the angle at an intermediate value of $h=0.3$ with reasonably high values of $\sqrt{\tau_3}\approx 0.1$. The independence of the latter on the alignment of the field is observed also for small values of $\gamma$ up to $\gamma=0.3$ with a value of $\sqrt{\tau_3}$ about $0.05$.
For models away from the symmetric XX model, we immediately see the typical maximum for $\sqrt{\tau_3}$ at vanishing angle $\alpha$. Away from this maximum $\sqrt{\tau_3}$ rapidly decays about an order of magnitude when the angle is about $\alpha=0.03 \,\pi$. 
This could be used to trigger the entanglement in a switch like situation having as a knob the orientation of the magnetic field. Likewise the independence of the entanglement of the field orientation might be used in environments where the orientation of the field is much less accurately controllable than its strength as a guaranteed source of a threetangled state.

We do not show curves for higher values of $\gamma$ and refer to a forthcoming publication\cite{longer-version}. However, the only behavior which survives there is the quick decay to about one order of magnitude from the maximum value for vanishing $\alpha$\cite{longer-version}. There is no reasonably large $\tau_3$ any longer.  
The region of strictly absent $\tau_3$ is slightly enlarged from $h=0.1$ at $\gamma=0.1$ to $h=0.3$ for $\gamma=0.5$ for being reduced again for larger values of $\gamma$. Systems with an anisotropy parameter of above $0.5$ could be still used as sensitive entanglement switch with the orientation of the magnetic field.
    
\begin{figure}[h!]
    \centering
    \includegraphics[width=0.5\linewidth]{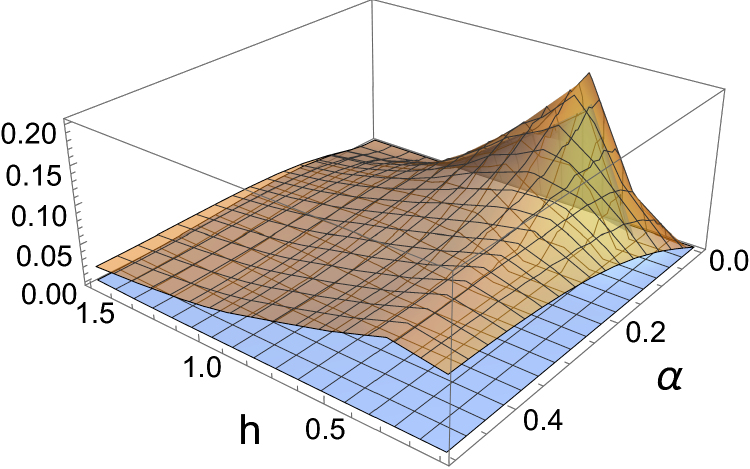}
    \includegraphics[width=0.48\linewidth]{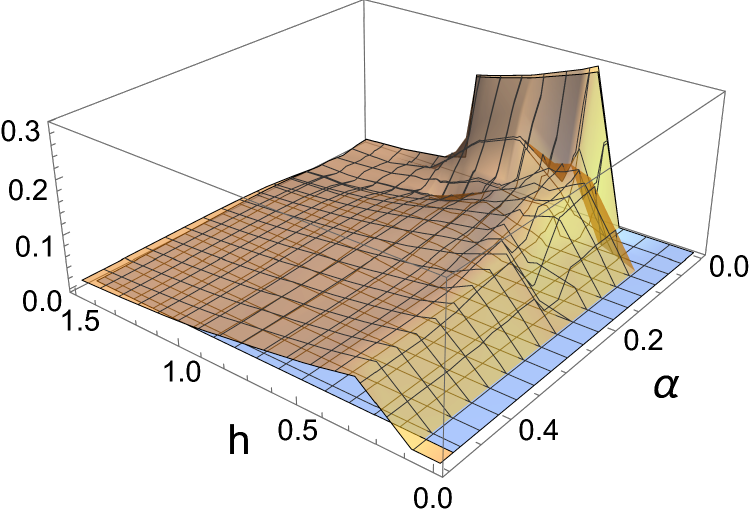}
    \caption{$\sqrt{\tau_3}$ as function of $\alpha$ and $h$ for the XX model (left) and the XY model with $\gamma=0.1$. Comparison of the result found here (transparent orange) with the lower bound from GHZ-symmetric states (blue). A reasonably large difference, in particular for values which are close to the isotropic XX model (left), is observed as already noticed in Ref.~\cite{osterloh2016exact}. }
    \label{fig:tau3-alpha-comparison}
\end{figure}

We compare the results with the lower bound from GHZ-symmetric states\cite{eltschka2012entanglement,siewert2012quantifying}
in Fig.~\ref{fig:tau3-alpha-comparison}.
For higher values of $\gamma$ and in the limit of $\alpha\to\pi/2$ it approximates our results well with more pronounced under-estimations for small values of $\alpha$ where our result turns to a provably exact value\cite{LOSU,KENNLINIE}. Inaccuracies in the algorithm are mainly due to the convexification of $(1,2)$ and $(2,2)$ decompositions, which would result in invisible changes on the scale of the tangle. These under-estimations of the lower bound have been first noticed and outlined in Ref.~\cite{osterloh2016exact}. For smaller values of $\gamma$ and in particular for the XX-model, it fails to predict the finite value of $\sqrt{\tau_3}$ also for higher values of $\alpha$ which could gain experimental relevance. 
There the model is apparently sufficiently far from the GHZ-symmetric case.

{\em Conclusions --}
The transverse XY models have been studied theoretically in a symmetry breaking field of given strength but with a deviation angle $\alpha$ from transversality. The model has been analysed for four sites, which provides a standard situation where the density matrix on three sites is strictly of rank 2 and therefore it is accessible to a quasi-exact treatment. The measure of entanglement is the genuine multipartite SL-entanglement, measured here by the square root of the threetangle. 

The general behavior is a detrimental effect whenever field inclinations are observed. Only the model with small anisotropy parameter $\gamma$ shows a remarkable resistance to this symmetry breaking field. Wheras the isotropic transverse model is disentangled in the threetangle it develops a steep maximum or pike at about $h=0.5$ and $\sqrt{\tau_3}\approx 0.2$. It tends to be independent of $\alpha$ for basically all investigated values of $h$ but with a maximum observed about $h=0.3$. In a reasonable range about this value of $h$, we observe a fixed value of $\sqrt{\tau_3}\approx 0.1$.
For an inhomogeneity of $\gamma=0.5$ and larger, this effect vanishes.
Hence, from an initially disentangled state the symmetry breaking and mixing of the two parity sectors of the $\ZZ_2$ symmetric model creates the entanglement for this case.
This is important from an experimental point of view because one can choose the parameter region properly and generates an output state with a reasonable value of $\sqrt{\tau_3}\approx 0.1$ which also is observed to be quasi-pure. This is not observed using the lower bound. The output state could then be fed into entanglement purification/distillation protocols. This would guarantee a source where this class of entanglement is necessary. 
The system could also be used as a sensitive entanglement switch around the maxima of the tangle at $\alpha=0$. It would be a highly sensitive detector of the magnetic field direction.
The tangle could be checked experimentally either by full state tomography as a proof of principle but also by every measure of multipartite entanglement\cite{zhang2023metrological,cao2024genuine,malla2023detecting,wu2024improved,li2024high} or  witnesses\cite{guhne2009entanglement,igloi2023entanglement}, once the nature of entanglement has been established. Also the detection via Machine Learning or Neural Networks, as suggested in Refs.~\cite{luo2023detecting,gulati2024ann} is a possibility.

As byproduct, a behavior of optimal decompositions for general SL-tangles could be singled out: simplified they are brachiating through states, which are either visible states (from the polar axis of the Bloch spere) of the zero-polytope on the one side and specific entangled states on the Bloch sphere on the opposite side. These states must be found by a convexification protocol.
The peculiar behavior of optimal decompositions could be harnessed for many important tasks as looking towards an order-dependent deviation as perturbation theory. For this task, it would be of utmost importance to develop lower bounds along the lines of Ref.~\cite{eltschka2012quantitative,eltschka2014practical}.

Finally, the quasi-pureness of the eigenstates apparently survives switching on the integrability breaking field. This could be used to calculate the three-tangle for these models along the lines of Ref.~\cite{osterloh2016upper}.

{\textit{Acknowledgements} --} We thank TII and in particular L. Amico for supporting this research.

\begin{appendix}


\section{Detecting genuine three-partite entanglement via the threetangle}\label{app:threetangle}

We will consider $\sqrt{|\tau_3|}$ as entanglement measure, where
the threetangle $|\tau_3|$ has been defined as\cite{Coffman00} 
(see also in Refs.~\cite{Wong00,VerstraeteDM03,OS04})
\nbeqa
\tau_3 &=& d_1 - 2d_2 + 4d_3  \\
  d_1&=& \psi^2_{000}\psi^2_{111} + \psi^2_{001}\psi^2_{110} + \psi^2_{010}\psi^2_{101}+ \psi^2_{100}\psi^2_{011} \\
  d_2&=& \psi_{000}\psi_{111}\psi_{011}\psi_{100} + \psi_{000}\psi_{111}\psi_{101}\psi_{010}\\ 
    &&+ \psi_{000}\psi_{111}\psi_{110}\psi_{001} + \psi_{011}\psi_{100}\psi_{101}\psi_{010}\\
    &&+ \psi_{011}\psi_{100}\psi_{110}\psi_{001} + \psi_{101}\psi_{010}\psi_{110}\psi_{001}\\
  d_3&=& \psi_{000}\psi_{110}\psi_{101}\psi_{011} + \psi_{111}\psi_{001}\psi_{010}\psi_{100}\ \ ,
\neeqa
and coincides with the three-qubit hyperdeterminant\cite{Cayley,Miyake02}. 
It is the only continuous SL-invariant here, 
meaning that every other such SL-invariant for three qubits can be expressed 
as a function of $\tau_3$.
It detects states from the only genuine SL-entangled GHZ-class.
$W$-states are not detected; they are instead detected as entangled by the pairwise Concurrence which is distributed along all possible pairs in the state. It is therefore not bipartite.

\section{The fourtangle vector}\label{app:fourtangle-vec}
Here we briefly review some aspects of Ref.~\cite{OS04,DoOs08}, with focus on the fourtangle.
The realm of entanglement families can be subdivided by the classification from Ref.~\cite{VerstraeteDMV02}, and the classification system of Ref.~\cite{OS04} with reference to filter operators.
These are antilinear operators that have zero expectation values on all possible bipartite pure states.
Three such filter operators have been pinpointed with whom a complete classification of the entanglement of four qubits is possible\footnote{The missing antilinear operator is necessary only to capture the group anti-symmetric under permutations of sites\cite{DoOs08}.}
and even for five qubits there is strong evidence that the system provided by local invariant operators (abbreviated by the word combs) be complete\cite{DoOs08}.
It is known that besides the trivial SL-invariant operator 
$H\propto\sigma_y^{\otimes 4}\cal{C}$ (with $\cal{C}$ the complex conjugation) there exist three operators as non-orthogonal basis to give every SL-invariant operator (see Ref.~\cite{DoOs08} and references therein).
One possible choice is ${\cal F}^{(4)}_{1}$, $\left\langle{\cal F}^{(4)}_2\right\rangle_s$, and ${\cal F}^{(4)}_{3}$, as defined in \cite{OS04,DoOs08}.
The corresponding SL-invariant quantities that quantify their entanglement are $\tau_{4a}[\psi]:=\bra{\psi}{\cal F}^{(4)}_{1}\ket{\psi}$, $\tau_{4B}[\psi]:=\bra{\psi}\left\langle{\cal F}^{(4)}_2\right\rangle_s\ket{\psi}$, and $\tau_{4c}[\psi]:=\bra{\psi}{\cal F}^{(4)}_{3}\ket{\psi}$.
In order for them to being comparable we look at their values which scales linearly with the density matrix, hence quadratically with the corresponding wave function, as the concurrence,
which results in $\vec{\tau}_4=(|(\tau_{4a})^{1/3}|,|(\tau_{4B})^{1/4}|,|(\tau_{4c})^{1/6}|))$ of length $\tau_4:=|\! |\vec{\tau}_4|\! |$. These values define the entanglement vector of a given state. Even though no direct analogy can be drawn with linear vector spaces, it has an important meaning concerning the entanglement classification in that a non-zero value is indicative for the entanglement content because the values of a given state will not change under SL transformations, which means a simple rescaling.
This gives rise to interesting transformation properties on given states\cite{EOSU,viehmann2012rescaling}. It would be intriguing to see the transformation properties of the optimal decompositions found here.

\section{General properties of optimal decompositions for real states of rank two}\label{app:optimal-decompositions}

In Refs.~\cite{caratheodory1911variabilitatsbereich,Uhlmann98} it is shown that for the number of pure states contained in any optimal decomposition, $n_{\rm opt}$, we have ${\rm rank} [\rho]
\leq n_{\rm opt} \leq ({\rm rank} [\rho])^2$. Here, it is thus sufficient to look for up to $4$ such pure states that generically form three-dimensional simplices.
We want to emphasize that more than the maximal $4$ states may be in an optimal decomposition; in this case, since every sub-partition of an optimal decomposition is itself optimal, each $4$ states out of that decomposition are optimal as well and the optimal decomposition is given by the convex polytope made out of these points. The tangle will behave linearly in these optimal polytopes.
If the density matrix is made of real entries only, as is the case for the ground state of a Hamiltonian which is non-degenerate, then every tangle $\tau$ has the property that $\tau(z)\equiv \tau(z^*)$ where $\ket{\psi}\propto\ket{\psi_1}+z \ket{\psi_2}$ is a representation that corresponds to a vector onto the Bloch-sphere with the parametrization\cite{osterloh2016exact}
\beq
z \mathrel{\widehat{=}}\vec{n}=\Matrix{c}{
x \\
y \\
z
}=\Matrix{c}{
2\sqrt{p(1-p)}\cos\phi \\
2\sqrt{p(1-p)}\sin\phi \\
2p-1
}
\eeq
where $\ket{\psi_i}$ are the two anti-podes on the z-axis.
Real superpositions lie in the $x$-$z$-plane.
If two vectors corresponding to $z_i$, $i=0,1$ yield a decomposition of $\rho$ on the $z$-axis with $z_0$ being on an external edge of the zero-polytope, so do
$z_i^*$. Therefore, if $\overline{z_1\, z_1^*}$ does not intersect an existing optimal polytope, the union of pure decomposition states is also a decomposition with the same tangle. Therefore, we can take the real values $z_{i,r}$ as reference values for searches after optimal decompositions.
This leads to 
\beq
z_{i,r}\mathrel{\widehat{=}}\vec{n}_i=\Matrix{c}{
2\sqrt{p_i(1-p_i)}\cos\phi_i \\
0\\
2p_i-1
}
\eeq
where it will be of particular interest a) where the line $\overline{\vec{n}_0\vec{n}_1}$ cuts the $z$-axis,  b) which are the weights of the corresponding entangled state(s), and c) where it crosses the Bloch sphere,
that is the corresponding pure states.
Whereas the answer to a) is given by the probability
\beq
P=\frac{p_0\sqrt{p_1(1-p_1)}\cos\phi_1+p_1\sqrt{p_0(1-p_0)}\cos\phi_0}{p_0\sqrt{p_0(1-p_0)}\cos\phi_0+p_1\sqrt{p_1(1-p_1)}\cos\phi_1}
\eeq
such that the point is located in $\vec{n}_P=(2P-1)\vec{e}_z$,
the answer to b) is given by 
\beq
\lambda=\frac{\mu_1}{\mu_0}=\frac{\sqrt{p_0(1-p_0)}\cos\phi_0}{\sqrt{p_1(1-p_1)}\cos\phi_1}
\eeq
and for the mainly interesting case of $\phi_1=0$ and $p$ being the probability in the density matrix in consideration, we obtain as a solution to c)
\begin{widetext}
\beq
P_\pm=p_0+\frac{p-p_0}{2}\frac{(p-p_0)(1-2 p_0)+2 p_0(1-p_0)\cos^2\phi_0\pm\sqrt{(p_0-p)^2+4 p p_0(1-p_0)(1-p)\cos^2\phi_0})}{(p-p_0)^2+ p_0(1-p_0) \cos^2 \phi_0}
\eeq
\end{widetext}
We refer to the following subsections for a collection of formulae used.

\begin{figure}[h!]
    \centering
    \includegraphics[width=\linewidth]{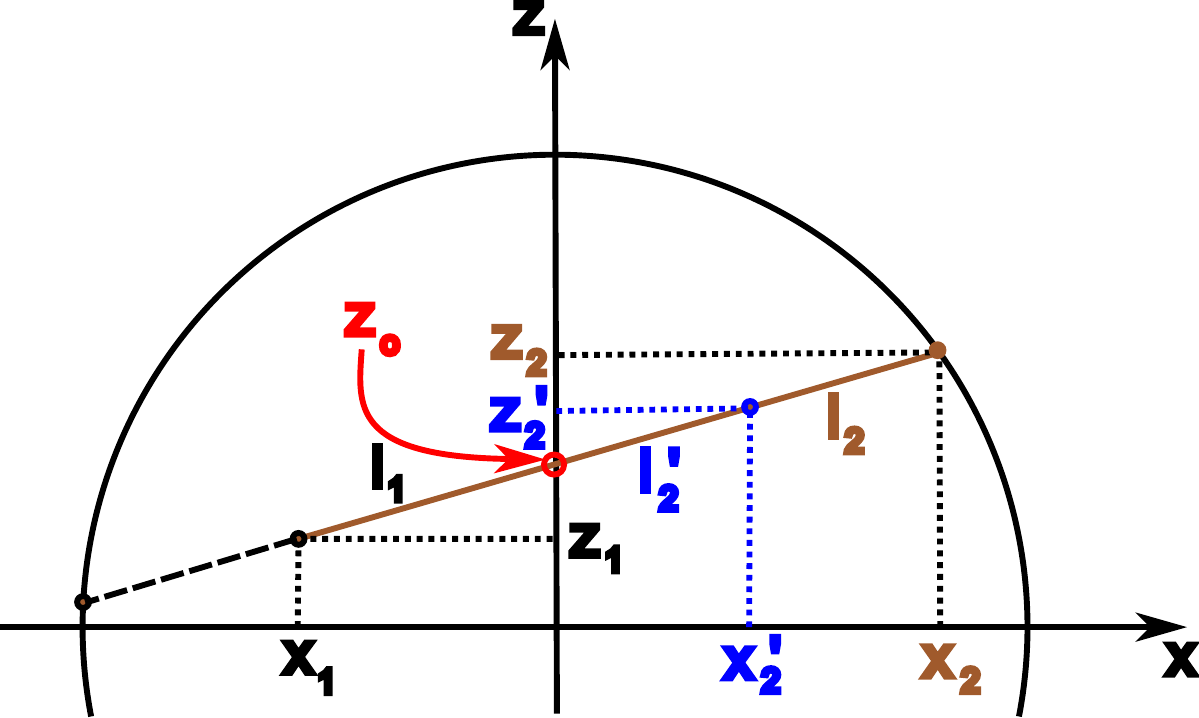}\hfill
   \caption{States within the Bloch sphere. For a given line $\overline{x_1x_2}$ all triangles $z_0x_1z_1$, $z_0x_2z_2$, and $z_0x_2'z_2'$, are similar. Ratios of their length are determined by the theorem of intersecting lines.}
    \label{fig:intersectinglines}
\end{figure}

The resulting tangle is a convex combination of the tangle values of the respective pure states.
This strictly linear behavior inside all the simplices of an optimal decomposition indirectly tells about whether there needs to be one or more states in a decomposition for being optimal:
the continuation of the respective decomposition type has to lead to a convex behavior in the tangle value; where this condition is not satisfied a (linear) convexification is needed. This holds in particular if the corresponding optimal polytope changes dimension. Here, this means to a dimensionality of at most three.
Next, we consider the weight laid on a state to ensure that the convex combination is in $2p-1$ on the $z$-axis,
where $p$ is the weight of the density matrix.
It must be noted that for $(n_0,n_e)$-decompositions with $n_e>1$, this weight on the $n_e$ pure states necessarily increases as the projection on the $x$-$z$-plane moves closer to the density matrix in consideration (see Fig.~\ref{fig:intersectinglines}. 
This must be overcompensated by a decreasing of the tangle of the corresponding pure states on the Bloch sphere.
Therefore, a steep relative maximum would be favorable for this to happen; however, far away from the states $\ket{Z_i}$, the tangle behaves smoothly and in particular, there are no ripples seen.
The presence of such ripples would indicate a much higher degree of the respective polynomial, which however would be reflected mainly in more zeros.
Except for the one point inside the Bloch sphere projection of Fig.\ref{fig:intersectinglines} being made out of three states from the
zero-polytope, the decompositions will be two-dimensional. 
If the curve of $(2,1)$-decompositions is strictly convex it need not to be looked after a vertical splitting of the pure entangled state into two (see however Fig.~\ref{fig:case-of-different-states}). The resulting $(2,2)$-decompositions would result in a three-dimensional tetrahedron which cuts the polar line of the Bloch sphere in an interval of non-zero length. This indicates a linear slope of the tangle. But every linear connection line lies above a strictly convex behavior of the tangle. Hence it cannot be optimal in this case.


\subsection{$(1,1)$-decompositions}

Here, we collect the formulae obtained for $(1,1)$-decompositions of a pure state at z-coordinate $2p_1-1$, where $p_1$ corresponds to its probability of the two states $\ket{\psi_i}$; $i\in\{N,S\}$. This corresponds to two pure states in the Fig.~\ref{fig:intersectinglines}.
We find
\beqa
l_1(p_1,p)&=&2\sqrt{(p-p_1)^2+p_1(p_1-1)}\\
l_2(p_1,p)&=&\frac{4p(1-p)}{l_1(p_1,p)}\\
m_2(p_1,p)&=&\frac{l_1(p_1,p)}{l_1(p_1,p)+l_2(p_1,p)}\\
p_2(p_1,p)&=&\frac{4p^2(1-p_1)}{l_1(p_1,p)^2}\\
&=&(1-p_1)\frac{p}{1-p}\cdot\frac{l_2(p_1,p)}{l_1(p_1,p)}\\
p(p1,p2)&=&p_1+\frac{(p_2-p_1)\sqrt{p_1(1-p_1)}}{\sqrt{p_1(1-p_1)}+\sqrt{p_2(1-p_2)}}
\eeqa
Next we want to examine whether for a given $(1,1)$ decomposition with $p_1$, $p_2$, and $p$ it is convenient to split the pure state into two, having a $(1,2)$ decomposition.
It can be seen that the ratios $x_1/(2|p_1-p|)=x_2/(2|p_2-p|)=x'_2/(2|p'_2-p|)$ are equal with $y'_2=\pm\sqrt{4p'_2(1-p'_2)-{x'_2}^2}$ (see Fig.~\ref{fig:intersectinglines}).
For the new $p'_2$ of this new state at phase $\phi_2$ the calculation gives
\begin{widetext}
\beq
p'_{2}(p_1,p;\phi_2)=p_1+(p-p_1)\frac{2p_1(1-p_1)-(p-p_1)(2p_1-1)\cos^2 \phi_2+\cos \phi_2\sqrt{4p(1-p)p_1(1-p_1)+(p-p_1)^2 \cos^2 \phi_2}}{2(p_1(1-p_1)+(p-p_1)^2\cos^2 \phi_2)}
\eeq
\end{widetext}

\subsection{$(2,2)$-decompositions}

In the following we will assume that two complex conjugated pairs, corresponding to a superposition of two states at $p_i$ and angles $\pm \phi_i$, $i\in\{1,2\}$, such that both states come to lie inside the Bloch sphere projection.
We find
\begin{widetext}
\beqa
m_2(p_1,\phi_1;p_2,\phi_2)&=&\frac{\sqrt{p_1(1-p_1)\cos^2(\phi_1)}}{\sqrt{p_1(1-p_1)\cos^2(\phi_1)}+\sqrt{p_2(1-p_2)\cos^2(\phi_2)}}\\
&=&
\frac{l_1(p_1,\phi_1;p_2,\phi_2;p)}{l_1(p_1,\phi_1;p_2,\phi_2;p)+l_2(p_1,\phi_1;p_2,\phi_2;p)}\\
m_1(p_1,\phi_1;p_2,\phi_2)&=& 1-m_2(p_1,\phi_1;p_2,\phi_2)\\
p(p_1,\phi_1;p_2,\phi_2)&=&p_1+(p_2-p_1)\frac{\sqrt{p_1(1-p_1)\cos^2(\phi_1)}}{\sqrt{p_1(1-p_1)\cos^2(\phi_1)}+\sqrt{p_2(1-p_2)\cos^2(\phi_2)}}\\ &=&\frac{p_2\sqrt{p_1(1-p_1)\cos^2(\phi_1)}+p_1\sqrt{p_2(1-p_2)\cos^2(\phi_2)}}{\sqrt{p_1(1-p_1)\cos^2(\phi_1)}+\sqrt{p_2(1-p_2)\cos^2(\phi_2)}}\\
&=& p_1 m_1(p_1,\phi_1;p_2,\phi_2)+p_2 m_2(p_1,\phi_1;p_2,\phi_2)\\
l_1(p_1,\phi_1;p_2,\phi_2;p)&=&2\sqrt{(p_1-p)^2+p_1(1-p_1)\cos^2 \phi_1}\\
l_2(p_1,\phi_1;p_2,\phi_2;p)&=&2\sqrt{(p_2-p)^2+p_2(1-p_2)\cos^2 \phi_2}
\eeqa
\end{widetext}
From this it is seen that the weights of the states are symmetric under exchanging $p_i\to 1-p_i$ separately for $i\in\{1,2\}$.

Usually, only one or two pure states are known (from the solutions for the zero-polytope) together with the parameter $p$ of the mixed state on the $z$-axis. A laborious way is 
calculating the respective variables via the theorem of implicit functions. We have not chosen this path and preferred to solve the convex and hence linear algebraic geometric equations instead which result from the vector form of the states. 
Setting $\phi_2=0$, the result is
\begin{widetext}
\beq
p_2(p_1,\phi_1;p)=p_1+(p-p_1)\frac{(p_1-p)(2p_1-1)+2p_1(1-p_1)\cos^2 \phi_1+\sqrt{(p_1-p)^2+4p(1-p)p_1(1-p_1)\cos^2 \phi_1}}{2((p_1-p)^2+p_1(1-p_1)\cos^2 \phi_1)}
\eeq
\end{widetext}
Next, for the evaluation  of $(2,2)$ from $(2,1)$ decompositions for given $p_1$, $\phi_1$, and $p$ the ratios of the following quantities are kept fixed (see Fig.~\ref{fig:intersectinglines}): $x_1/(2|p_1-p|)=x_2/(2|p_2-p|)=x'_2/(2|p'_2-p|)$ with $y'_2=\pm\sqrt{4p'_2(1-p'_2)-{x'}_2^2}$. These have to be substituted in $m_2(p_1,\phi_1;p'_2,\phi'_2)$ where $x'_2=2\sqrt{p'_2(1-p'_2)}\cos \phi'_2$.

\section{Case studies for the $3^+N$  an $3^-N$ scenario}\label{app:case-studies-3pmN}

Here we analyze exemplarily the state emerging as ground state for the values $h=0.5$, $\gamma=0.1$ and $\alpha=0.15708$ with a four tangle vector of $\vec{\tau}_4 \approx (0.0802,0.0648,0.0934)$ with length $\tau_4\approx 0.1391$ in case of the $3^+N$ class and $\vec{\tau}_4\approx (0.0273,0.0393,0.0484)$ with length $\tau_4=0.0680$ in case of the $3^-N$ class.
Fig.~\ref{fig:(3,1)-decomposition} shows  
$\sqrt{\tau_3}$ and the $(2,1)$ (red) and $(1,1)$ (blue and purple) decompositions for this specific case. The minimum is given by the black curve underneath. It is clearly seen that a convexification is needed. 
\begin{figure}[h!]
    \centering
    \includegraphics[width=0.45\linewidth]{Figs/_2,1_-_1,1_-Decompositions-plot-h-0-5-gamma-0-1-alpha-0-1.pdf}\hfill
    \includegraphics[width=0.45\linewidth]{Figs/3-1-simplices-2-1inmidst.pdf}
    \includegraphics[width=0.45\linewidth]{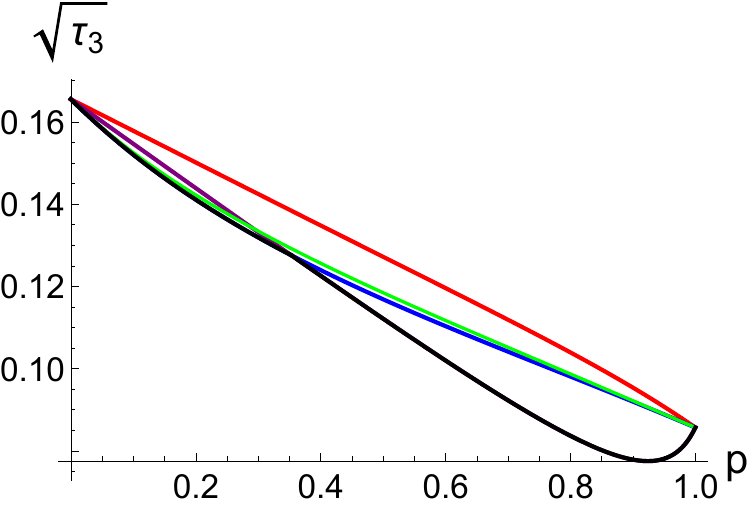}\hfill
    \includegraphics[width=0.45\linewidth]{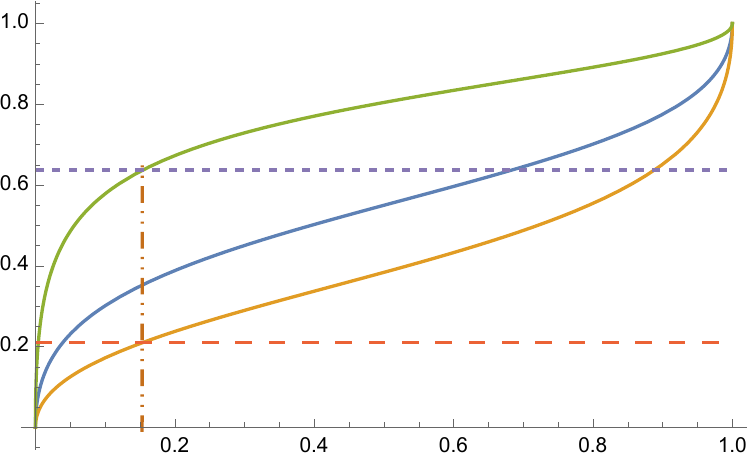}
   \caption{Left panel: $\sqrt{\tau_3}$ is shown for the $(2,1)$  (red) and $(1,1)$ (blue and purple) decompositions for a zero-polytope of classes $3^+N$ (upper row) and $3^-N$ (lower row). The minimum is given by the black curve underneath. 
   Right panel: the piercing point of the polar axis of the Bloch sphere is plotted as a function of the $p$-value of the corresponding pure state as tip of that tetrahedron.
   See also Fig.~\ref{fig:case-of-different-states}.}
    \label{fig:(3,1)-decomposition}
\end{figure}
It leads for the $3^+N$ class to a single 3D simplex on top of each visible sides of the zero-polytope as a basis. This can be seen in the right panel of the plot where the piercing point of the central axis is plotted as a function of the $p$-value of the corresponding pure state as tip of that tetrahedron.
For the class $3+N$, single pure states at $p$-values of $p_{N^+}\approx 0.8692$ and $p_{N^-}\approx 0.1296$ can be singled out. In the $3^-N$ class it is a single triangular tip at $p_{\Delta}\approx 0.1539$. Also here multiple states in the optimal decomposition would be indicated by multiple $p$-values for them in the right corresponding figure. Such a situation would be supported by an additional maximum in the $(1,1)$ and $(2,1)$ decompositions as sketched in Fig.~\ref{fig:case-of-different-states}.

We have analysed whether deviations in $\phi$ direction are possible to lower the overall value for $\sqrt{\tau_3}$ with respect to $(1,1)$ decompositions. It is shown a light green curve which corresponds to $(1,2)$ decompositions with a yet considerable value of $\phi=0.5$. The growing of the weight function slightly overcompensates the
gain one would have in the threetangle of the corresponding pure states.
As argued earlier, a splitting in $p$ direction would immediately lead to linearization in a 3D simplex; this would lead to a linearization in an already strictly convex background and hence to a bigger value for $\sqrt{\tau_3}$.
\begin{figure}[h!]
    \centering
    \includegraphics[width=\linewidth]{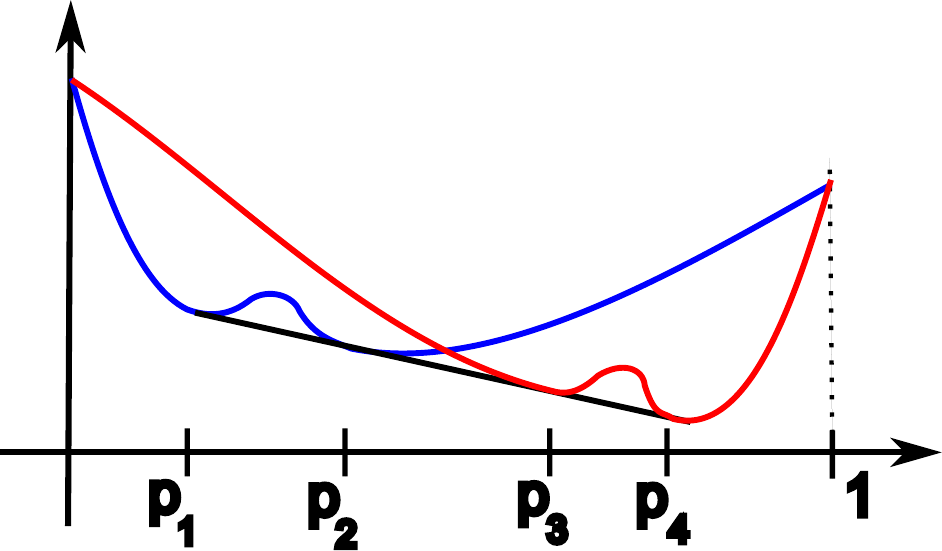}
    \caption{If the states would not coincide in the procedure described above this would mean that the corresponding $(1,1)$ and $(2,1)$ decompositions (red and blue) should possess a relative maximum as described.}
    \label{fig:case-of-different-states}
\end{figure}

Summarizing for the case $3^+N$, this convexification leads to a single 3D simplex on top of each visible sides of the zero-polytope as a basis which cut the central $z$-axis from  $p_{\rm low}^-\approx 0.2487$ to $p_{\rm high}^-\approx 0.4476$ and from $p_{\rm low}^+\approx 0.7913$ to $p_{\rm high}^+\approx 0.9141$. This can be seen in the corresponding right panel of the plot where the piercing point of the central axis is plotted as a function of the $p$-value of the corresponding pure state as tip of that simplex.
Indeed, single pure states at $p$-values of $p_{N^+}\approx 0.8692$ and $p_{N^-}\approx 0.1296$ can be singled out.
Multiple pure states would be indicated by different corresponding values of $p$ for them. In this case, however, both the corresponding $(1,1)$ and $(2,1)$ decompositions should have a relative maximum, as sketched in Fig.~\ref{fig:case-of-different-states},
which is not the case here.

The same procedure for a zero polytope of the $3^-N$ type leads to the curves plotted in the lower two plots in Fig.~\ref{fig:(3,1)-decomposition}.
Again, the left figure shows the curves for $(2,1)$ (red) and $(1,1)$ (blue and purple) decompositions, whereas on the right the convexification points are shown with $p^-\approx 0.2098$ and $p^+\approx 0.6368$. Both points coincide in a single pure state at $p_\Delta\approx 0.1539$.

\section{Case studies for the $4Y$ scenario}\label{app:case-studies-4Y}

\begin{figure}[h!]
    \centering
    \includegraphics[width=0.45\linewidth]{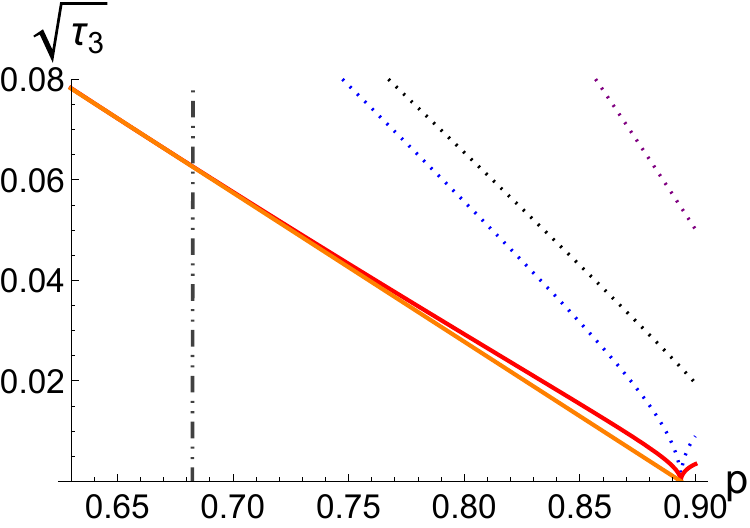}
    \includegraphics[width=0.45\linewidth]{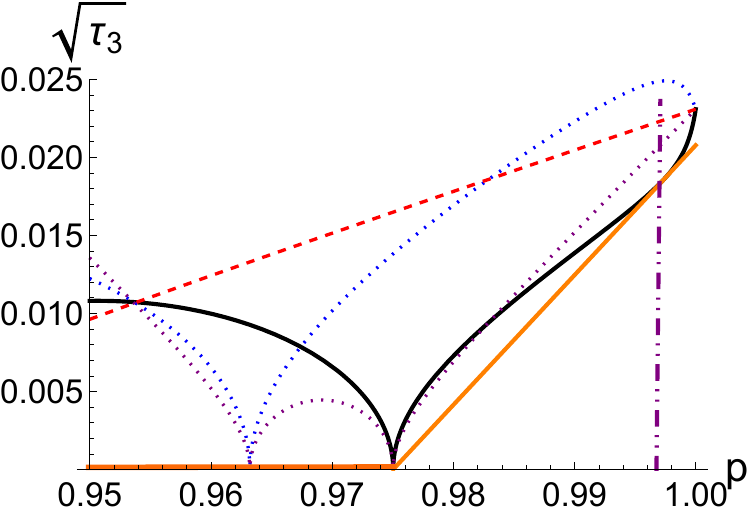}
    \includegraphics[width=0.45\linewidth]{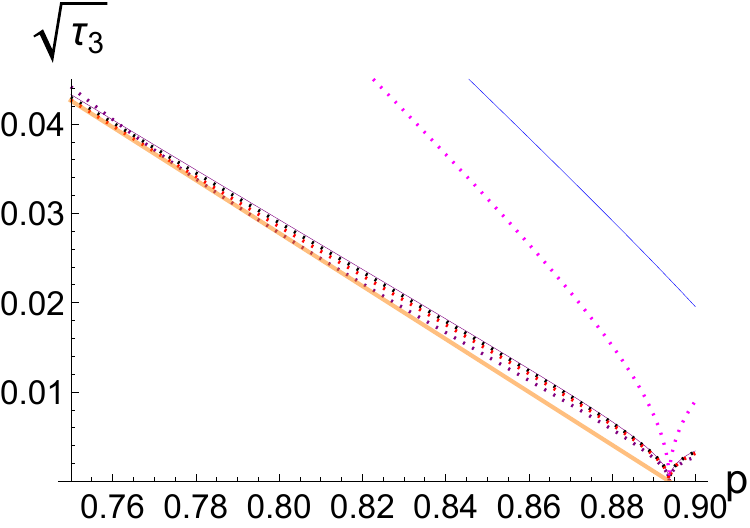}
    \includegraphics[width=0.45\linewidth]{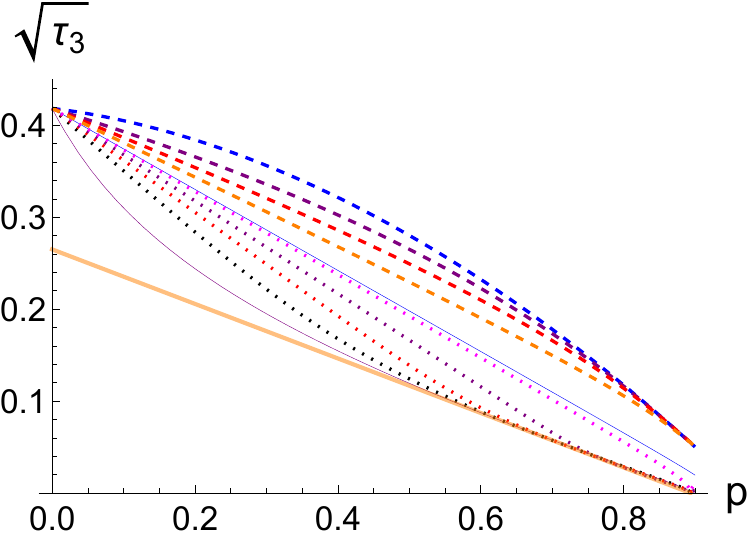}
    \caption{Convexification procedure in the case $4Y$, where four different real solutions, both negative and positive, are present in the model. Here is shown a particular case for $\gamma=1$, $h=0.9$ and $\alpha=0.1 \pi$, but this case is generic for this particular case of $\gamma=1$. The convexification of the curve leads to the position of $\ket{N_\pm}$.}
    \label{fig:case-study-fourReals-convexification}
\end{figure}

A case study of this scenario is shown in Fig.~\ref{fig:case-study-fourReals-convexification} where the zero-states $\ket{Z_i}$ are situated as in Fig.~\ref{fig:zero-polytope-class} and the number of the state is in decreasing order of $p$. $(1,1)$ decompositions with all zero-states $\ket{Z_{3,4}}$ are shown. 
Here is shown a particular case for $\gamma=1$, $h=0.9$ and $\alpha=0.1 \pi$, but this case is generic for this particular case of $\gamma=1$. The solutions $Z_i$ of the zero polytope (zero-states) are located at values $p_i=0.999142$, $0.998093$ (both solutions have positive sign), $0.567217$ , and $0.119002$ (both with negative sign); the respective tangle vanishes in between $0.893712$ and $0.975046$ (see also~\cite{osterloh2016exact}). The particular convexification points are indicated by dash-dot-dotted vertical lines at $p_{-}=0.656507$ and $p_{+}=0.997357$ in the first line of figures. Shown are the 1-1-decompositions with the first (black) to fourth (red) zero-state $\ket{Z_i}$ and their convexifications in orange. 
In the second row of figures (2,1)-decompositions are shown (dotted lines) where convex combinations of $\ket{Z_2}$ and $\ket{Z_4}$ are the disentangled part of the decomposition; they come to lie all above the convexification.
In the cases we have inspected, the weight of the corresponding eigenstate was located within the strictly convex zones. We have therefore not implemented the convexification procedure into the algorithm.
\end{appendix}

\bibliography{bib,biblio}

\begin{thebibliography}{70}
\bibitem [{\citenamefont {Osterloh}()}]{longer-version}%
  \BibitemOpen
  \bibfield  {author} {\bibinfo {author} {\bibfnamefont {A.}~\bibnamefont
  {Osterloh}},\ }\href@noop {} {\enquote {\bibinfo {title} {More details on the
  threetangle in the xy-model class with a non-integrable field background},}\
  }\bibinfo {note} {To be published soon}\BibitemShut {NoStop}%
\end{thebibliography}

\end{document}